\documentclass[journal]{IEEEtran}
\usepackage[noend]{algpseudocode}
\usepackage{algorithmicx,algorithm}
\usepackage{graphicx}
\usepackage{amsmath}
\usepackage{amssymb}
\usepackage[caption=false]{subfig}
\usepackage[noadjust]{cite}
\usepackage{float}
\usepackage{color}
\usepackage{url}
\usepackage{comment}

\definecolor{armygreen}{rgb}{0.0, 0.5, 0.0}

\begin{document}

\title{Noisy Pooled PCR for Virus Testing}

\author{
Junan Zhu, Kristina Rivera, and Dror Baron\textit{ Senior Member, IEEE}\thanks{
The work of Baron was supported in part by NSF EECS $\#1611112$. Rivera was supported by the National Institutes of Health under award number F31DK118859-02 and by the 2019 Howard Hughes Medical Institute Gilliam Fellowship Award. 

Zhu is with Harvest Fund Management, Beijing, China, email junan.zhu@gmail.com.
Rivera and Baron are with North Carolina State University,
Raleigh, NC, email $\{$krrivera, barondror$\}$@ncsu.edu. 
}}
\date{\vspace{-5ex}}
\maketitle

\begin{abstract}
Fast testing can help mitigate the coronavirus disease 2019 (COVID-19) pandemic.
Despite  their  accuracy  for  single  sample  analysis, infectious diseases diagnostic tools, like
RT-PCR,  require  substantial  resources  to  test  large  populations.
We develop a 
scalable approach for determining the viral status of 
pooled patient samples. Our approach converts group
testing to a linear inverse problem, where 
false positives and negatives are interpreted as generated by a noisy communication channel, and a message passing algorithm estimates the illness status of patients.
Numerical results reveal that our approach estimates patient illness 
using fewer pooled measurements than existing noisy group testing algorithms. 
Our approach can easily be extended to various applications, including where false negatives must be minimized. Finally,
in a Utopian world we would have collaborated with RT-PCR experts; it is difficult to form such connections during a pandemic.
{\em We welcome new collaborators to reach out and help improve this work!}
\end{abstract}


\section{Introduction}\label{sec:intro}

{\bf Motivation.}
{\em Reverse transcription polymerase chain reaction} (RT-PCR) is a prevalent diagnostic tool for infectious diseases, 
like {\em coronavirus disease 2019} (COVID-19). 
RT-PCR is a labor intensive technical procedure that requires
numerous trained laboratory personnel to analyze one patient sample~\cite{Nolan2006}. 
Briefly, {\em ribonucleic acid} (RNA) 
is isolated from a patient's respiratory tract and purified for reverse transcription, a process where the RNA template is turned into {\em complementary deoxyribonucleic acid} (cDNA). cDNA, along with specific viral primers, is loaded into a machine, where cDNA is 
amplified and annealed to the target sequence. While extending through each PCR cycle, a reporter dye 
is cleaved or broken from a probe to amplify fluorescence intensity and reveal a positive 
sample.

Despite its accuracy for 
single sample analysis, RT-PCR requires substantial resources to test a large number of samples. Instead, we aim to develop a
scalable testing procedure that allows for patient samples to be combined before PCR.

{\bf Main idea.}
Noisy group testing is used to analyze RT-PCR data from mixed or {\em pooled} samples, as recently demonstrated for COVID-19~\cite{Kishony2020}. 
The goal of group testing is twofold. First, to increase the accuracy of
testing for each individual patient by combining information
from multiple pooled measurements that sample genetic material from that same individual.
Second, to use a reduced number of measurements,
especially in settings where a large population is being tested, most patients
are healthy, and so many individual measurements will come out negative
and thus show that multiple patients are healthy. In summary, group testing allows to evaluate
large populations at high throughput, low per-patient diagnostic costs,
and low false positive and negative probabilities.

Noiseless group testing has been established, but noisy group testing algorithms are less mature. For example, recent work on COVID-19 \cite{Hanel2020,Kishony2020}
uses pooled tests to rule out patients corresponding to negative pooled measurements.
Their approach implicitly relies on false negatives being rare in RT-PCR,
but diluting many samples may increase false negatives~\cite{Stramer2013}. 
Additionally, patients corresponding to 
positive pooled measurements are later tested individually~\cite{Hanel2020},
which does not benefit from pooling.
Our algorithm (Sec.~\ref{sec:GAMP}) applies pooling to identify individual 
sick patients.

Recently, researchers across the world have been looking to increase the sample size per PCR run by using 
custom barcodes for each sample and then pooling them together \cite{PennState2020}. 
Custom barcoding is not new in terms of multiplexed genetic sequencing \cite{Baym2015}. 
Briefly,  custom  barcodes  for  each  patient’s RNA sample are designed by an
algorithm  and  substituted  in  as the  {\em reverse  transcriptase}  (RT)  primers  to  generate  barcoded cDNA. 
Next-generation sequencing is performed after a single pooled PCR reaction, and then  demultiplexed  to  
determine  each  sample’s  viral  content. Our method of
pooling samples before adding barcodes  could be 
used by researchers for a quicker time to analysis and also as a complementary method to reanalyze barcoded samples.

\begin{figure}[!t]
\centering
\includegraphics[width=0.4\textwidth]{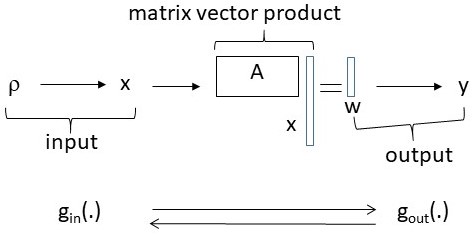}
\caption{System model. 
A Bernoulli process with probability $\rho$ of sickness 
generates an input vector $x\in \{0,1\}^N$, which reflects patient illness status.
The input is multiplied by a measurement matrix, $A\in \{0, 1\}^{M\times N}$,
resulting in noiseless measurements, $w=Ax \in \mathbb{N}^M$ (\ref{eq:w}),
which are processed by an RT-PCR channel, resulting in noisy 
measurements, $y\in \{0,1\}^M$ (\ref{eq:channel}). GAMP~\cite{Rangan11}
processes the input channel relating $\rho$ and $x$ with $g_{in}(\cdot)$ (\ref{eq:g_in}),
and the output channel relating $w$ and $x$ with $g_{out}(\cdot)$ (\ref{eq:g_out})
(details in Sec.~\ref{sec:GAMP}).}
\label{fig:model}
\vspace{0.in}
\end{figure}

{\bf Contributions and organization.}
We focus on a simple pooled testing model for RT-PCR (Sec.~\ref{sec:model}).
This model is converted to a linear inverse problem (Sec.~\ref{sec:noiseless}),
and our goal is to estimate a vector of patient illness status, $x$,
from a vector of noisy RT-PCR measurements, $y$, 
a matrix $A$ relating patients and measurements, 
and statistical information about false positives and negatives (Sec.~\ref{sec:noisy}
and Fig.~\ref{fig:model}).
This estimation problem is solved using 
{\em generalized approximate message passing} (GAMP) \cite{Rangan11} in Sec.~\ref{sec:GAMP}.
Promising numerical results are provided in Sec.~\ref{sec:numerical},
and Sec.~\ref{sec:discussion} discusses how our GAMP-based approach can be extended.

\section{Model}
\label{sec:model}

{\bf Conventional RT-PCR.}
RT-PCR has a binary outcome. That is, once the sample is amplified,
there is plenty of genetic material available for identification. 
Prior to amplification, there can be problems during pre-processing to isolate purified RNA.
Either genetic material can be damaged, in which case all further tests with this material
are negative, or the sample is contaminated, in which case further tests are positive.
However, such pre-processing problems essentially flip the sample's condition permanently from negative to positive or vice versa, 
because all tests of the patient
will be using flawed genetic material
from this patient. Therefore, such problems will not be discussed further.

Once the sample has been pre-processed, there are two further problems that could
arise once we partition the sample into multiple group test measurements. 
One possible outcome is a {\em false negative}, meaning 
not enough genetic material from a sick patient and, therefore, insufficient amplification. 
It is also possible to have a {\em false positive}, meaning
the sample was contaminated by viral matter, and the test is positive although the patient is healthy. 
We focus on these two problems, as a well-designed group testing procedure mitigates their effects.

{\bf Group testing.}
Instead of sampling genetic material from one patient, we will pool
material from multiple patients. In principle, if any of the patients is
sick, and there is sufficient genetic material, then the pooled group 
test will come out positive. However, it is possible that group testing will use
less genetic material per patient, meaning that the measurement is diluted, 
and the probability of false negatives (per sick patient) might be larger
than conventional (unpooled) measurements~\cite{Stramer2013}.

\subsection{Number of sick patients per measurement}
\label{sec:noiseless}

We now express the number of sick patients pooled per measurement
as a product between a binary matrix representing what patients are pooled in different measurements, 
and a binary vector representing patient illness status.
Later, Sec.~\ref{sec:noisy} forms a noisy probabilistic outcome, where the RT-PCR
test being positive or negative depends probabilistically on 
the number of sick patients pooled per measurement.

Our system is illustrated in Fig.~\ref{fig:model}. 
We have $N$ patients.
The status of each patient is given by 
$x_n$, where $n \in \{1, \ldots, N\}$.
If patient~$n$ is sick, then $x_n=1$, else $x_n=0$.
The $N$ entries are modeled as
{\em independent and identically distributed} (i.i.d.),
and we model $x_n$ using a {\em random variable} (RV), $X_n$.
These $N$ RVs follow
a Bernoulli {\em probability mass function} (pmf), 
$X_n \sim \text{Ber}(\rho)$,
\[
\Pr(X_n=1)=\rho \quad \text{and} \quad \Pr(X_n=0)=1-\rho,
\]
where $\Pr(\cdot)$ denotes probability, and $\rho$ is the percentage of sick patients.

Next, the vector $x \in \{0,1\}^N$ is multiplied 
by a binary {\em measurement matrix}~$A\in \{0, 1\}^{M\times N}$.
Because $x$ and $A$ are binary, the matrix vector product,
\begin{equation}
\label{eq:w}
w=Ax\in\mathbb{N}^M,
\end{equation}
is a length-$M$ vector of natural (non-negative) numbers.

The matrix $A$ is interpreted as follows.
Row~$m$ corresponds to measurement~$m$, and
column~$n$ to patient~$n$, where
$m\in\{1,\ldots, M\}$ and $n\in\{1,\ldots, N\}$.
If patient~$n$ is not measured in measurement~$m$,
then $A_{mn}$, the matrix entry in row~$m$ and column~$n$, is zero;
such patients do not affect the outcome of measurement~$m$.
In contrast, $A_{mn}=1$ when genetic material from patient~$n$
appears in measurement~$m$. 
It can be seen that $w_m$ 
counts the number of sick patients evaluated by measurement~$m$. 

In noiseless group testing,
the RT-PCR measurement is positive if and only if $w_m > 0$.
However, RT-PCR suffers from false positives and negatives.

\subsection{Noisy model}
\label{sec:noisy}

We now account for these false positives and negatives.
The noisy measurement $y_m$ depends on $w_m$ through a conditional probability,
$\Pr(Y_m|W_m)$, where $Y_m$ and $W_m$ are RVs.
To evaluate $\Pr(Y_m|W_m)$, we
denote the probability of an individual patient being sick
yet not having enough genetic material in one of the measurements by $p_1$.
This is the probability of a false negative caused by {\em one} patient;
with $w_m$ sick patients, the probability that all of them
have false negatives is $(p_1)^{w_m}$.
(We note in passing that false negatives could be modeled
as independent of $w_m$~\cite{Hanel2020}.)

Similarly, the probability of a false positive is denoted by $p_2$.
If there is a false positive in $y_m$, then $y_m=1$, irrespective
of the status of the patients being evaluated by measurement~$m$.
On the other hand, $y_m=0$ means that there was no false positive,
and all the patients evaluated that were actually sick resulted in false negatives.
Based on this discussion, we can express the probability for $y_m$ to be
0 or 1 given $w_m$,
\begin{equation}
\label{eq:channel}
\Pr(Y_m=0|W_m=w_m) = (1-p_2)(p_1)^{w_m}.
\end{equation}
Then, we compute $\Pr(Y_m=1|W_m)=1-\Pr(Y_m=0|W_m)$.
In communication and information theory, such a
probabilistic relationship is known as a {\em channel}~\cite{Cover06};
the output channel relating the vectors
$w$ and $y$ appears in Fig.~\ref{fig:model}.

We now have a linear relationship from $x$ to $w$,
and the noiseless measurements vector $w$, which contains the
number of sick patients per measurement,
is then processed by a probabilistic channel to yield the
noisy measurements vector, $y$.
Our goal is to estimate
$x$ from $y$, $A$, and statistical information about the channel.
Other group testing approaches often perform pooled measurements in
a first part, and positives are tested individually in a second part~\cite{Hanel2020};
our method can improve both parts by pooling all measurements and accounting
for all available information.
In the following section, we describe our algorithmic framework in detail.

\section{Algorithmic framework}
\label{sec:GAMP}

We estimate $x$ from $y$, $A$, and statistical information about the channel
by applying {\em generalized approximate messsage passing} (GAMP)~\cite{Rangan11},
which is an iterative signal estimation algorithm.
GAMP is preferred, because it achieves best-possible estimation-theoretic performance 
asymptotically, in the limit of large linear estimation problems. 

Our approach focuses on the
{\em large system limit}, where $N \rightarrow \infty$,
$M(N)$ depends on $N$, 
and $\lim_{N \rightarrow \infty}\frac{M(N)}{N}=R$,
where we call $R$ the measurement rate.
GAMP relies on the large system limit
for various summations in the derivation steps of the algorithm to be well-approximated
as Gaussian under the central limit theorem~\cite{Rangan11}. 
Note that running our algorithm for small problem sizes such as $N=100$
patients and $M=30$ measurements may result in poor estimation quality. 

The GAMP algorithm is listed in Algorithm~\ref{algo:gamp}. For a detailed derivation, 
we refer the reader to Rangan~\cite{Rangan11}.
An intuitive and less formal explanation is provided below.

\begin{algorithm}[t]
\caption{GAMP}\label{algo:gamp}
{\bf Inputs.} Maximum iterations $t_{max}$,
percentage of sick patients $\rho$, false negative probability $p_1$, false positive probability $p_2$, measurements $y$,
and matrix $A$. \\
{\bf Initialize.} 
$t,k,h_m,\Theta_m,\widehat{x}_n,s_n, \forall m,n$.\\
{\bf Comment.} 
$t$ is iteration number,
$k$ is mean of our estimate for $Ax$,
$h_m$ is correction term for $w_m$,
$\Theta_m$ is variance of $h$,
$\widehat{x}_n$ is our estimate for $x_n$, 
$s_n$ is variance in our estimate $\widehat{x}_n$.
\begin{algorithmic}[1]
\While{$t < t_{max}$} 
\State // {\bf clean up output channel}
\State $\Theta=(A)^2s$ // variance of $h$
\State $k=A\widehat{x} - \Theta h$ // mean of $w$ per previous iteration
\For{$m=1$ to $M$}
\State $h_m= g_{out}(k_m,y_m,\theta_m)$
\State {\bf Comment:} $\frac{1}{\Theta}(E[W_m|K_m,Y_m,\Theta_m]-k_m)$. 
\State $r_m= -\frac{\partial}{\partial k_m} g_{out}(\cdot)$
\EndFor
\State $\Delta_v = \left\{ \frac{1}{N} (A^T)^2 r \right\}^{-1}$ // scalar channel noise variance
\State $q = \widehat{x} + \Delta_v A^T h$ // pseudo data \label{line:q}
\State // {\bf clean up input channel}
\For{$n=1$ to $N$}
\State $\widehat{x}_n = g_{in}(\Delta_{vn},q_n) = E[x_n|q_n]$ // mean estimate 
\State $s_n = E[x_n^2|q_n] - E^2[x_n|q_n]$  // variance estimate
\EndFor
\State $t=t+1$
\EndWhile
\end{algorithmic}
{\bf Output.} Estimate $\widehat{x}$, pseudo data $q$,
and scalar channel noise variance $\Delta_v$.
\end{algorithm}

GAMP is comprised of two parts. The first part involves the input channel 
relating $\rho$ and $x$ (Fig.~\ref{fig:model})~\cite{Rangan11}, 
where $x$ is estimated from an
auxiliary vector $q \in \mathbb{R}^N$ (cf. Line~\ref{line:q} of Algorithm~\ref{algo:gamp}) through a function $g_{in}(\cdot)$. 
The auxiliary vector, $q$, is known in the AMP literature\footnote{AMP can be derived from GAMP for a specific setting~\cite{Rangan11}. 
While AMP~\cite{DMM2009} requires the matrices it processes to have zero mean, GAMP is 
less restrictive.} 
as the pseudo data, which can be treated as a noisy version of the true signal $x$,
\begin{equation}
\label{eq:q}
q=x+v,
\end{equation}
where $v \in \mathbb{R}^N$ is {\em additive white Gaussian noise} (AWGN)
with zero mean, where $\Delta_{vn}$ is the variance of $v_n$.
Hence, $g_{in}(\cdot)$ can be interpreted as a denoising function,
\begin{equation}
\label{eq:g_in}
\widehat{x}_n = g_{in}(\Delta_{vn},q) =  E[X_n|Q_n=X_n+{\cal{N}}(0,\Delta_{vn})].
\end{equation}
While other denoising functions can be used, conditional expectation,
i.e., $E[X|Q]$, minimizes the {\em mean squared error} (MSE)
in each GAMP iteration, and so it reduces the error as quickly as possible.

The second part of GAMP involves the output channel (cf. Fig.~\ref{fig:model}), 
where $y_m$ depends probabilistically on $w_m$.
We estimate $w_m$ from $y_m$ using a second denoising function,
\begin{equation}
\label{eq:g_out}
h_m = g_{out}(k_m,y_m,\Theta_m) = \frac{E[W_m|K_m,Y_m,\Theta_m]-k_m}{\Theta},
\end{equation}
where the expectation is taken over the pmf,
\[
f(w_m|k_m,y_m,\Theta_m) \propto \Pr(y_m|w_m)\exp\left[-\frac{(w_m-k_m)^2}{2\Theta_m}\right].
\]
In this expression, (\ref{eq:g_out}), we have mean and variance values for $w_m$,
and can interpret $g_{out}(k_m,y_m,\Theta_m)$ as 
a correction term that reflects residual information,
which is provided by the noisy measurements
vector, $y$, but is not yet reflected in our estimates,
$k$ for $Ax$, and $\widehat{x}$.
The correction term is used in later iterations
to compute $q$ and $g_{in}(\cdot)$. 

GAMP uses these two scalar functions, 
$g_{in}(\cdot)$ (\ref{eq:g_in}) and $g_{out}(\cdot)$ (\ref{eq:g_out}), 
to estimate $x$ and $w=Ax$ (\ref{eq:w}) from 
$q$ (\ref{eq:q}) and $y$ (\ref{eq:channel}), 
respectively. That is, GAMP iteratively cleans the input and output channels. 
A numerical illustration is provided in Fig.~\ref{fig:channel_evolution}; Sec.~\ref{sec:numerical} 
discusses this figure in detail. GAMP also uses derivatives of these scalar functions to estimate
the variance. In words, knowing not only the mean but also the variance around the mean
allows GAMP to judiciously use information from $\widehat{x}$ when estimating $\widehat{w}$ and vice versa.

\section{Numerical Results}\label{sec:numerical}

\subsection{GAMP illustration} 

This section provides numerical results showing how GAMP solves noisy group testing problem.
For readers who are new to GAMP, we begin by illustrating how GAMP cleans up the input and output channels 
iteratively.

We evaluate $N=5000$ patients at a time, where the fraction of infected patients is $\rho=0.01$. 
The measurement rate is $R=M/N=0.3$, meaning that we take $M=NR=1500$ RT-PCR measurements.\footnote{
Our GAMP-based algorithm is relatively fast; problems of size $(M=1500, N=5000)$
take a few seconds to run on a laptop computer.}
The matrix $A$ is designed to pick up $n_{pos}$
sick patients per measurement on average; we let $n_{pos}=0.5$. The numbers of ones per row and column are kept close to 
$n_{pos}/\rho$ and $Rn_{pos}/\rho$, respectively.
For the RT-PCR channel, we assume a false negative probability, $p_1=0.02$, 
and false positive probability, 
$p_2=0.001$.\footnote{
The parameters $p_1$ and $p_2$ resemble Hanel and Thurner~\cite{Hanel2020};
other sources suggest larger false positive and negative probabilities.
For our software, these are merely parameters that are easily modified.}
We quantify GAMP signal estimation quality using the {\em area under the receiver operating curve} (AUC-ROC).
In words, the ROC captures trade-offs between false positives and negatives,
and increasing the AUC reflects better estimation.
While standard GAMP minimizes the MSE~\cite{Rangan11},
other error metrics can be minimized~\cite{Tan2014,ZhuBaron2018}.

\begin{figure}[!t]
\centering
\includegraphics[width=0.5\textwidth]{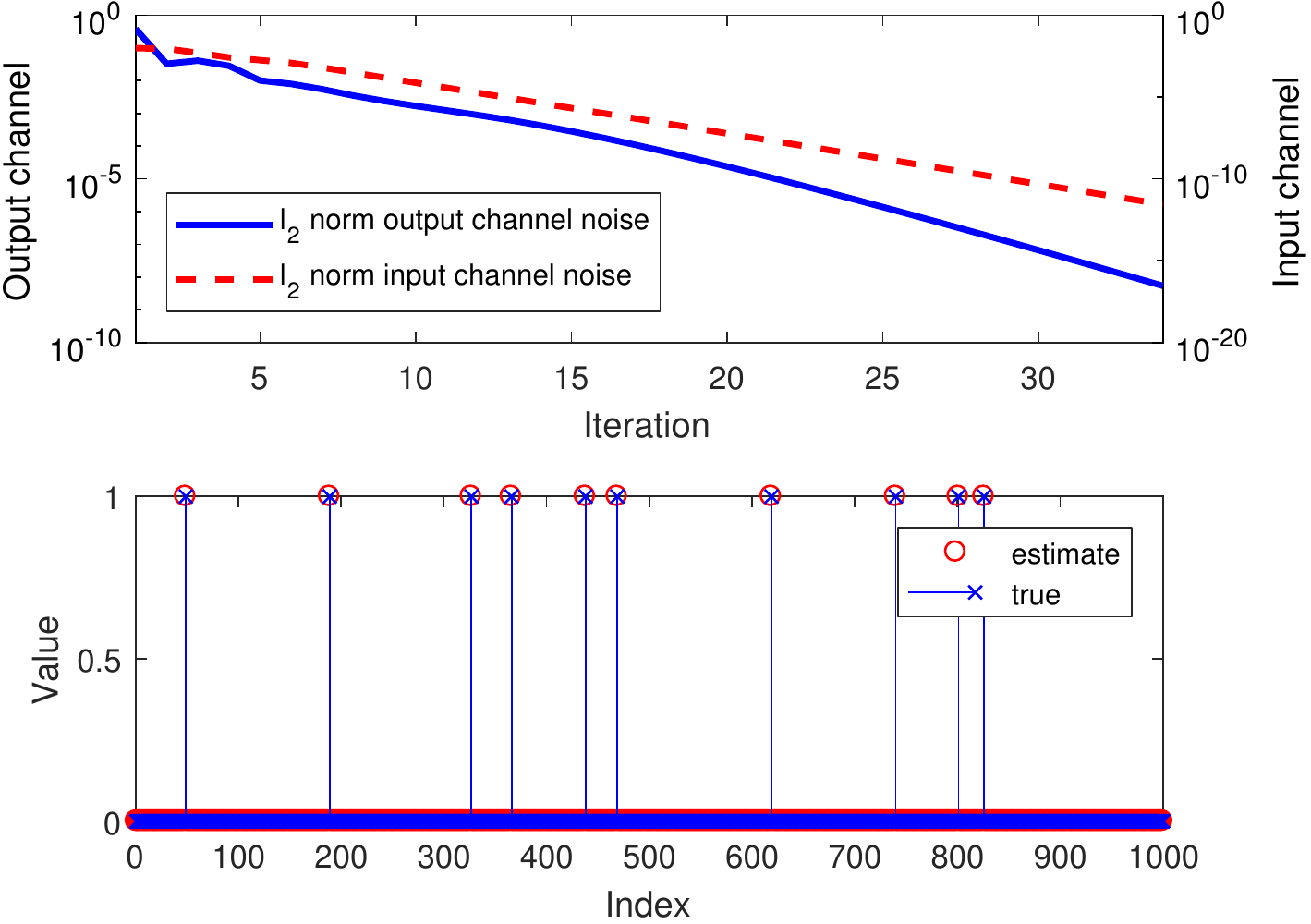}
\caption{Top: $\ell_2$ norms of the input 
(dashed red line; associated with right vertical axis) and 
output (solid blue; left) channel noise as functions of the GAMP iteration. 
Bottom: The first 1000 entries of the unknown patient illness status
vector $x$, and their estimates.
($N=5000$ patients; 
$\rho=0.01$ percentage of sick patients;
measurement rate $R=M/N=0.3$; 
$n_{pos}=0.5$ average sick patients per measurement; 
false negative probability $p_1=0.02$; 
false positive $p_2=0.001$.)}
\label{fig:channel_evolution}
\vspace{0.in}
\end{figure}

The top pannel of Fig.~\ref{fig:channel_evolution} plots the $\ell_2$ norms of the input channel noise (dashed red line) 
and output channel noise (solid blue). We can see that 
the input and output channels improve over iterations. 
The bottom panel shows the first 1000 entries of the input signal vector $x$ and their estimates. 
We can see that patient illness status is estimated well.

\subsection{Group testing under various conditions}

\begin{figure}[!t]
\centering
\includegraphics[width=0.5\textwidth]{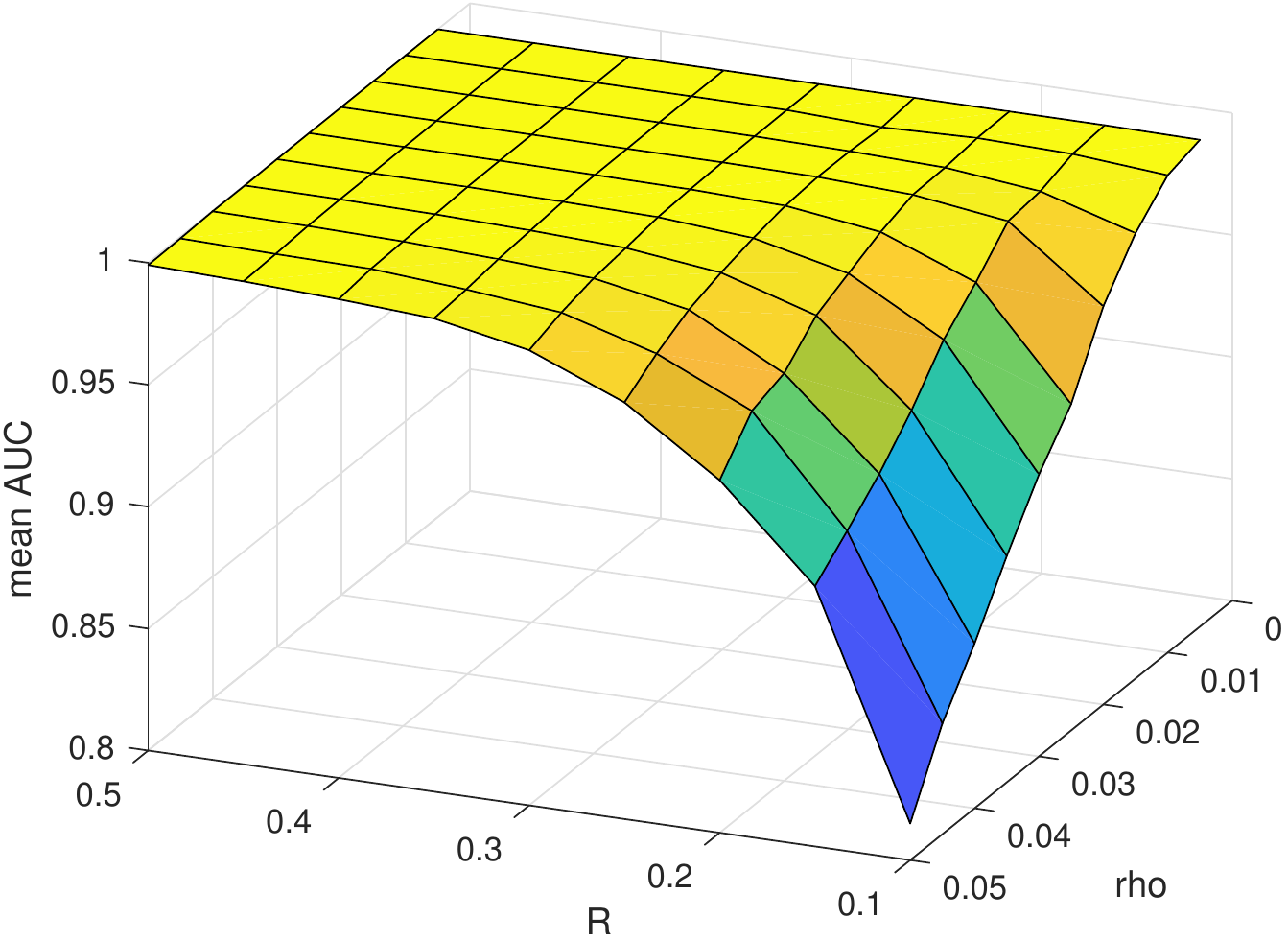}
\caption{
Estimation accuracy in AUC (vertical axis) as a function of 
the percentage of sick patients, $\rho$, and measurement rate, $R=M/N$.
($N=5000$; $n_{pos}=0.5$; $p_1=0.02$; $p_2=0.001$.)
}
\label{fig:R_vs_rho}
\vspace{0.in}
\end{figure}

We investigate the impact of the percentage of sick patients, $\rho$,
and measurement rate, $R=M/N$, on estimation accuracy in Fig.~\ref{fig:R_vs_rho}.
As before, 
$N=5000$, $n_{pos}=0.5$, $p_1=0.02$, and $p_2=0.001$. 
We run our algorithm on $\rho \in \{0.005, 0.01, 0.015, \cdots, 0.05\}$ 
and $R \in \{0.1, 0.15, 0.2, \cdots, 0.5\}$. 
For each setting, we randomly generate 20 different triples of $(x, A, y)$, 
and record the AUC for every triple. 
The performance for each setting is evaluated by averaging AUC values. 
Our results show that the AUC increases with the measurement rate, $R$, 
and larger $\rho$ requires larger $R$ to yield an AUC near 1.
These results align with our expectation that more measurements 
improve estimation, while more sick patients require more measurements.

\subsection{Two part approach}

Recent work Hanel and Thurner~\cite{Hanel2020}
analyzes a two part group testing approach.
Their model for PCR uses $\Pr(Y_m=0|W_m>0) =(1-p_2)p_1$, 
while we use $\Pr(Y_m=0|W_m=w_m>0) =(1-p_2)p_1^{w_m}$;
we evaluated their approach using $\rho=0.01$, $p_1=0.02$, and $p_2=0.001$.
Hanel and Thurner's Part~1 pools a block of $B=11$
patients at a time.
If a pool is negative, all patients in the block are declared healthy;
else Part~2 measures them individually.
The measurement rate is
\[
R=\frac{1}{B}+\Pr(\text{pool tests positive})= \frac{1}{B}+\Pr(Y_m=1).
\]
We compute $\Pr(Y_m=1)$,
\begin{eqnarray*}
\Pr(Y_m=1) 
&=& \Pr(W_m=0)\Pr(Y_m=1|W_m=0) \\
&+& \Pr(W_m>0)\Pr(Y_m=1|W_m>0).
\end{eqnarray*}
Note that
$\Pr(W_m=0)=(1-\rho)^B$,
$\Pr(Y_m=1|W_m=0)=p_2$,
$\Pr(W_m>0)=1-\Pr(W_m=0)$, and
\[
\Pr(Y_m=1|W_m>0)=1-(1-p_2)p_1.
\]
Combining these results,
\[
R=\frac{1}{11}
+0.99^{11} 0.001 + (1-0.99^{11})(1-0.999 \cdot 0.02) =0.1944.
\]
A simulation over $N=10^7$ patients had
935 false positives and 4038 false negatives.

Our two part approach modifies Part~2.
Instead of testing patients within each positive block individually,
we combine all patients within all positive blocks into a
new linear inverse problem, and solve the resulting estimation problem~\eqref{eq:w} 
with GAMP. For example, let the block size be $B=25$ patients in Part~1.
In Part~2, we combine all positive blocks and apply $R=0.5$ and $n_{pos}=0.5$ to the linear inverse problem.
Note that ({\em i}) positive measurements from Part~1 
are reused in the matrix $A$ and measurement vector $y$ of Part~2,
because they contain information that helps GAMP;
({\em ii}) we decide whether a patient is sick or not 
by thresholding $\widehat{x}$.
Combining Parts~1 and~2, the measurement rate is $R=0.149$.
We randomly generate 100 $(x,y,A)$ triples for 
$N=10000$ patients in Part~1,
Among $100N=10^6$ patients, there are
92 false positives and 366 false negatives.
Our false positive and negative rates are both
lower than those of Hanel and Thurner~\cite{Hanel2020}.

\section{Discussion}
\label{sec:discussion}

Our current approach relies on various assumptions. 
Below are issues that can be considered in ongoing and future work.

{\bf Challenges.} 
Some of the challenges we expect involve better modeling of RT-PCR,
in particular how pooling multiple samples dilutes the genetic material
and may increase false positives and negatives~\cite{Chatel2018}.
Other challenges involve matrix design; better matrices
will improve estimation quality.

One question is whether $p_1$ is the same for each patient~$n$ in each measurement~$m$ where
$A_{mn}=1$.
It might be possible to use different matrix entry values
(not just 0 and 1) and thus sample more genetic material in some
cases, and less in others. This will likely result in $p_1$
depending on the amount of genetic matter being sampled.
Therefore, genetic material will have to be measured before samples 
are pooled to ensure the same concentration of genetic material is loaded for each sample.
Limiting the amount of genetic matter being processed may 
reduce the costs of the overall measurement process. On the other hand, if the amount of genetic material per patient per measurement
varies, a sophisticated channel could be supported by having nonzeros in $A_{mn}$ take different values.

Another question is whether the false positive probability, $p_2$,
is identical for all measurements. Alternately, $p_2$ may depend on
the measurement system, for example the number of samples pooled together, 
or the number of RT-PCR iterations.
If individual RT-PCR iterations are costly, then the
number of iterations can be reduced, resulting in larger $p_2$.
Our experience with AMP-based algorithms suggests that a modest increase in $R=M/N$
will compensate for the degraded individual measurements. 
Cost effectiveness of each iteration will determine the number of times each sample can be run. 
This value should be weighed against the number of times individual samples are pooled, allowing to optimize the number of samples pooled per measurement \`a la~\cite{Hanel2020}.

A third challenge pertains to the measurement matrix, $A$.
In our current design, rows and columns contain similar numbers of nonzeros
(Sec.~\ref{sec:numerical}).
Therefore, each measurement provides the same {\em signal to noise ratio} (SNR).
One matrix design option is to allow different rows and columns to have different numbers of nonzeros.
Another is to prevent any pair of patients, $n_1$ and $n_2$, 
from both having nonzero matrix entries in different rows, $m_1$ and $m_2$, i.e.,
$A_{m_1n_1}$,
$A_{m_1n_2}$,
$A_{m_2n_1}$, and
$A_{m_2n_2}$ cannot all be nonzero. Refinements in matrix design will improve our estimation quality.

{\bf Applications.} Improvements in testing accuracy can be used in different ways in different applications.

\noindent \textbullet 
False positive and negative rates of individual RT-PCR measurements 
can be reduced by pooling together samples, and using GAMP-based
algorithms. This can help decide when it is safe to release a COVID-19 patient from quarantine.

\noindent \textbullet 
Latency reduction. As the first RT-PCR measurements from a batch of patients arrive, 
all patients corresponding to positive measurements can be quarantined.
As more RT-PCR measurements come in, GAMP can determine which 
of the individual patients in the positive pooled samples are actually healthy. 

\noindent \textbullet 
Throughput can be drastically increased for fixed target levels of false positives 
and negatives. This can be useful for testing large populations with minimal cost.

\noindent \textbullet 
False negatives must be low to prevent a few sick patients from infecting many others. Low false negative probabilities can be provided by a 
two-part signal estimation approach~\cite{MaBaronNeedell2014}. 
Part~1 will use a conventional measurement matrix $A$.
Part~2 takes extra measurements only for patients deemed healthy in Part~1,
thus reducing false negatives.
Similar ideas have been proposed for adaptive sensing~\cite{HN11}.

Finally, our GAMP-based approach uses statistical information about the
probability of a patient being sick, $\rho$, and probabilities governing false
positives and negatives, $p_1$ and $p_2$. The algorithm can be improved using more information,
for example statistical dependencies between household members 
being sick. We will integrate more predictive 
information to further improve estimation quality.

\section{Acknowledgments}

The authors thank Michael Daniele for RT-PCR modeling discussions, and 
Ahmad Beirami,
Yitzhak (Tsahi) Birk, 
Igor Carron, 
Steven Cotten,
Florent Krzakala, 
John Muth,
and Lenka Zdeborova for discussions of group testing. 
Baron also thanks numerous colleagues at North Carolina State University for putting him in touch with various specialists.

\bibliographystyle{IEEEtran}


\end{document}